\documentclass[12pt,preprint]{aastex}

\newcommand{\fer}{{\it Fermi}}
\newcommand{\wse}{{\it WISE}}
\newcommand{\strip}{{\it WISE blazar Strip}}
\newcommand{\ver}{{\it VER J0648+152}}
\usepackage{graphicx}
\usepackage{longtable}

\slugcomment{version \today: fm}

\shorttitle{Identification of the infrared non-thermal emission in blazars}
\shortauthors{F. Massaro, R. D'Abrusco, M. Ajello, J. E. Grindlay \& H. A. Smith  2011}

\begin{document}
\title{Identification of the infrared non-thermal emission in Blazars}
\author{F. Massaro\altaffilmark{1,2}, R. D'Abrusco\altaffilmark{2}, M. Ajello\altaffilmark{1}, J. E. Grindlay\altaffilmark{2} \& Howard A. Smith\altaffilmark{2}}

\affil{SLAC National Laboratory and Kavli Institute for Particle Astrophysics and Cosmology, 2575 Sand Hill Road, Menlo Park, CA 94025}
\affil{Harvard - Smithsonian Astrophysical Observatory, 60 Garden Street, Cambridge, MA 02138}

\begin{abstract}
Blazars constitute the most interesting and enigmatic class of extragalactic $\gamma$-ray sources
dominated by non-thermal emission.
In this Letter, we show how the \wse\ infrared data make possible to identify a distinct region of the 
[3.4]-[4.6]-[12] $\mu$m color-color diagram where the sources dominated by the the thermal radiation
are separated from those dominated by non-thermal emission, in particular the blazar population.
This infrared non-thermal region delineated as the \strip\ (WBS), 
will constitute a new powerful diagnostic tool when the full 
\wse\ survey data is released. The WBS can be used to extract new blazar candidates, 
to identify those of uncertain type and also to search for the counterparts of unidentified 
$\gamma$-ray sources.
We show one example of the value of the use of the WBS identifying the TeV source \ver, 
recently discovered by VERITAS.
\end{abstract}


\section{Introduction}\label{sec:intro}
Blazars are one of the most enigmatic and rare class of active galactic nuclei (AGNs). 
Their continuum emission is dominated by non-thermal radiation
from radio to $\gamma$-ray energies, making them the most
frequently detected class of extragalactic sources at GeV-TeV energies.
Their observational features also include high and variable polarization, 
superluminal motions, very high observed luminosities coupled with a flat radio spectrum 
that steepens in the IR-optical bands and a rapid variability from the radio to X-ray bands 
with weak or absent emission lines.
In 1978, Blandford and Rees suggested that radiation of blazars 
can be described as arising from a relativistic jet closely aligned to the line of sight \citep{blandford78}.
In short, blazars constitute the purest non-thermal kind of AGNs. 

The spectral energy distributions (SEDs) of blazars include two main 
components: a low-energy component with 
power peaking in the range from the IR to the X-ray band, 
and a substantial high-energy component often dominated by $\gamma$-rays.
Blazars come in two flavors: the BL Lac objects (BZBs) and Flat Spectrum Radio Quasars (BZQs),
with the latter having strong emission lines, generally higher radio to optical polarization
higher redshift and more prominent $\gamma$-ray bump of their SEDs, than the former population.
The BL Lac population is in turn divided in two subclasses: 
the ``low-frequency peaked BL Lacs"  (LBLs) 
in which the peak of the first component
falls in the IR-optical range, and the ``high-frequency peaked BL Lacs" (HBLs) 
when it falls in the range between the UV and the X-ray bands \citep{padovani95}.
Recently,  HBLs detected at TeV energies have been reclassified 
as TBLs \citep{massaro08,massaro11a}.

In this Letter we present a color-color diagram using infrared (IR) magnitudes
that allows us to distinguish between extragalactic sources dominated 
by non-thermal emission, like blazars, and other classes of  galaxies
and/or AGNs.
We construct the diagram 
using the archival data of the recent Wide Infrared Survey Explorer (\wse) facility
\citep{wright10}.
The \wse\ mission mapped the sky at 3.4, 4.6, 12, and 22 $\mu$m 
in 2010 with an angular resolution of 6$^{\prime\prime}$.1, 6$^{\prime\prime}$.4, 6$^{\prime\prime}$.5 
\& 12$^{\prime\prime}$.0 in the four bands, achieving 5$\sigma$ 
point source sensitivities of 0.08, 0.11, 1 and 6 mJy in unconfused regions on the ecliptic, 
respectively. Regarding the \wse\ astrometry we note that
the absolute (radial) differences between \wse\ source-peaks and ``true" astrometric positions 
anywhere on the sky are no larger than $\sim$ 0$^{\prime\prime}$.50, 0$^{\prime\prime}$.26, 0$^{\prime\prime}$.26, and 1$^{\prime\prime}$.4 for the
four \wse\ bands, respectively \citep{cutri11}
\footnote{{\underline http://wise2.ipac.caltech.edu/docs/release/prelim/expsup/sec2\_3g.html}}.

We show how in the [3.4]-[4.6]-[12]$\mu$m color-color diagram
the blazar population, which is dominated by non-thermal emission in the infrared, 
covers a distinct region well separated from the locus of
other extragalactic sources dominated by the IR thermal radiation.
We suggest how this infrared non-thermal region (hereinafter the \strip\,, WBS) 
could be used as a diagnostic tool to identify the blazar population
over the entire sky when the final data release of the \wse\ survey will be available.
Finally, we discuss the possible implications of our results on efforts to  
identify $\gamma$-ray sources and blazar of uncertain type.
As an example, we applied the WBS as a new diagnostic tool based on the \strip\ to 
look for the counterpart of the recently 
unidentified BL Lac candidate discovered by VERITAS: \ver\ \citep{ong10}.
We note that the \wse\ 22$\mu$m band also offers a similarly useful color diagnostic,
but because of its larger beam it is less useful in this investigation.

\section{Blazar \wse\ detections}
\label{sec:detect}
We considered all the blazars present in the ROMA-BZCAT
\footnote{{\underline http://www.asdc.asi.it/bzcat/}} \citep{massaro09,massaro10},
adopting the same nomenclature, namely: BL Lac objects (BZBs), 
flat spectrum radio quasars (BZQs), 
and blazar of uncertain type (BZUs).
We note that the class BZU which contains sources showing 
occasional presence/absence of broad spectral lines
or transition objects between a radio galaxy and a
BL Lac, as well as sources with limited spectral 
data does not allow a precise classification.

The total number of blazars in the ROMA-BZCAT that fall in the area surveyed by 
\wse\ during the first year  
(corresponding to 57\% of the whole sky) is 1487 (due to the inhomogeneity of the 
sky coverage of the ROMA-BZCAT). 
To search for the positional coincidences of blazars in the observed \wse\ sky
we considered two different regions.
The first of radius 2$^{\prime\prime}$.4
corresponding to the combination of the error of 1$^{\prime\prime}$ assumed for
the radio position reported in the ROMA-BZCAT \citep{massaro09}
and that of the fourth \wse\ band at 22$\mu$m (i.e., 1$^{\prime\prime}$.4)
(see Section~\ref{sec:intro} and reference therein);
The second region has a radius of 6$^{\prime\prime}$.5,
equal to the angular resolution of the third \wse\ band.
In our analysis, we excluded multiple crossmatches.
We also note that the cross-match between the ROMA-BZCAT and the \wse\ catalog
has been performed considering only the sources present in the first year \wse\ catalog detected 
with a minimum signal-to-noise ratio (SNR) higher than 7 in at least one band.

The number of positional coincidences within the first region of 2$^{\prime\prime}$.4 is 1365
corresponding to 92\% of the blazars in the \wse\ first year survey, detected
with a chance probability of 2.5\%, evaluated applying the same method 
described in Maselli et al. (2011). We did not find any multiple matches adopting the area of 2.4$^{\prime\prime}$.
The number of blazars associated with \wse\ sources increases to 1446 with a chance probability of 11\%
considering the second region of 6$^{\prime\prime}$.5, where we found 8 multiple matches.
Thus, 97\% of the blazars appear to have a counterpart in the \wse\ catalog.
To be more conservative in our analysis, we consider, for the rest of the discussion,
only those blazars with \wse\  sources associated within the 2$^{\prime\prime}$.4, 
unless stated otherwise.

\section{The \strip\ }
\label{sec:plot}
In the \wse\ Preliminary Source Catalog (WPSC) the
[3.4]-[4.6]-[12] $\mu$m color-color diagram, drawn from high and low galactic latitude regions, 
shows the location of different classes of objects \citep{cutri11}
\footnote{{\underline http://wise2.ipac.caltech.edu/docs/release/prelim/preview.html}}.
\begin{figure}
\includegraphics[height=8.cm,width=8.5cm,angle=0]{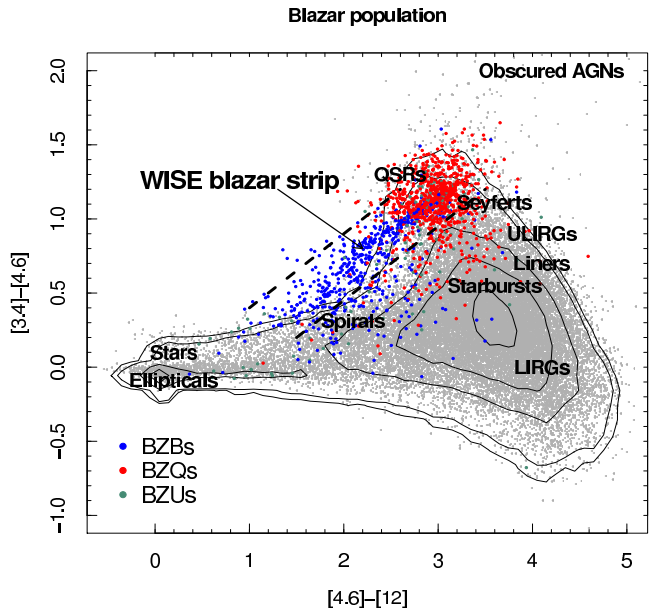}
\caption{The [3.4]-[4.6]-[12] $\mu$m color-color diagram of \wse\ thermal sources and blazars.
We report the 1365 blazars associated with a \wse\ source within a region of radius 2$^{\prime\prime}$.4.
The two blazar classes of BZBs (blue filled circles) and BZQs (red filled circles) are shown
together with the blazars of uncertain type (BZUs, green filled circles). The background grey dots
correspond to the 453420 \wse\ thermal sources detected in a region of 56 deg$^2$ at high Galactic latitude.
The isodensity curves for the \wse\ thermal sources, corresponding to 50, 100, 500, 2000 sources 
in the 56 deg$^2$ area are reported (see Section~\ref{sec:plot}).
The location of different classes of objects is also shown, where QSRs, ULRIGs and LIRGs
indicate the quasars, the ultraluminous infrared galaxies and the luminous infrared galaxies, respectively.
The WBS is highlighted within the two black dashed lines.}
\label{fig:ints_all}
\end{figure}

In order to put blazar populations on the same plot,
we selected 14 random regions of 4 deg$^2$ each, not overlapping,
for a total 56 deg$^2$ at high Galactic latitude, 
within the 116 deg$^2$ considered in the WSPC \citep{cutri11}.
Then, we collected all the 453420 sources detected by \wse\ in its 1st year catalog 
(hereinafter \wse\ thermal sources, because they are dominated by thermal emission in the infrared),
within the selected 56 deg$^2$ region, having SNR$>$7 at least one band,
a conservative level for the WPSC release to emphasize Catalog reliability$^{~3}$ \citep{cutri11}.
Being far from the Galactic plane, the majority of these sources have extragalactic origin
and only few stars lie in the selected region.

We built [3.4]-[4.6]-[12] $\mu$m color-color diagram
using the magnitudes reported in the \wse\ Catalog\footnote{All WISE magnitudes are in the Vega system.} 
for all the sources present in the 56 deg$^2$
and for all the blazars detected by \wse\ .
We note that the relative errors for both the infrared colors are less than 0.1
for 95\% of the entire blazar sample but less to 0.05 for $\sim$ 85\%.
In Figure~\ref{fig:ints_all}, the two subclasses of blazars and those of uncertain type are shown,
overlaid to the isodensity contours for all the sources in the 56 deg$^2$ of the \wse\ selected area.
The four levels of the isodensity contours correspond to
50, 100, 500, 2000 sources, in the 56 deg$^2$ area, respectively.
Finally, in Figure~\ref{fig:ints_all}, we also annotated the location of different classes of objects.

The main striking result is that the blazars lie in a distinct region (i.e., the WBS) of the 
[3.4]-[4.6]-[12] $\mu$m color-color plot with respect to the \wse\ thermal sources, in particular
where the density of the other extragalactic sources is decreasing 
that of the blazar population increases (see below for more details). 
Moreover the two main classes of BZQs and BZBs are themselves
separated in two distinct regions of the WBS, where BZQs are generally redder and
lie closer to the normal quasars (QSRs) and Seyfert galaxies.

\begin{figure}[!b]
\includegraphics[height=8.cm,width=8.5cm,angle=0]{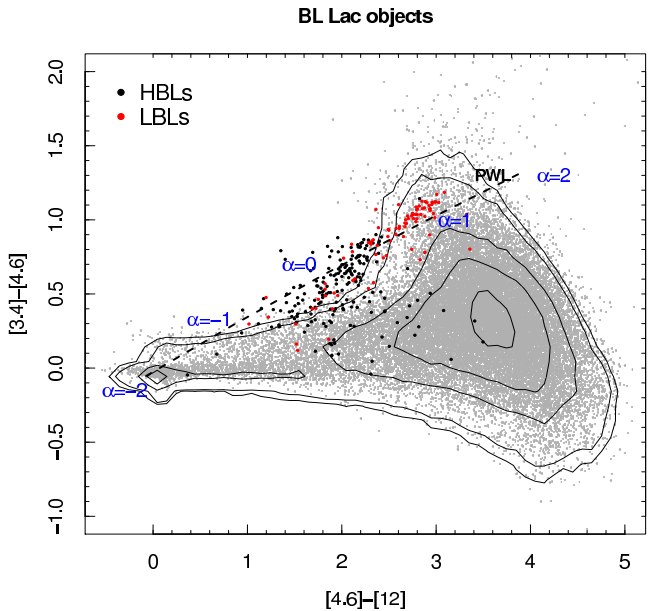}
\caption{Same of Figure~\ref{fig:ints_all}, only for the BL Lac objects. 
The two subclasses of HBLs (black circles) and LBLs (red circles) are highlighted. 
We also report the line PWL (black) correspondent to an IR non-thermal power-law spectrum
of spectral index $\alpha$. We marked the IR colors of different $\alpha$ values in blue
along the black dashed lines.
(see Section~\ref{sec:plot} for more details).}
\label{fig:ints_bzb}
\end{figure}

In Figure~\ref{fig:ints_bzb}, we consider only the BZBs
distinguishing between their subclasses of HBLs and LBLs. 
We calculated the ratio between the X-ray and the radio flux $\Phi_{XR}$
for those source with these fluxes available in the ROMA-BZCAT 
to distinguish between HBLs and LBLs in the above sample.
We adopted a very conservative classification considering as HBLs 
the BZBs with $\Phi_{XR}$ $\geq$ 1 \citep{maselli10} and LBLs for all the others. 

All the BZBs without X-ray data available are omitted from Figure~\ref{fig:ints_bzb}
because it is not possible to classify them. 
The BZUs have been excluded from our investigation, 
because of the uncertainties in their nature \citep{massaro09,massaro10}.

As many authors have pointed out the infrared spectrum
of BL Lac objects, being dominated by non-thermal emission, can be described as
a power-law with the spectral index $\alpha$ \citep[e.g.][]{impey82,bersanelli92}
{\bf ($S_{\nu} \propto \nu^{-\alpha}$).}
Then, different values of $\alpha$ correspond to different colors in the 
[3.4]-[4.6]-[12] $\mu$m color-color diagram \citep{wright10}.
In Figure~\ref{fig:ints_bzb}, we show the line corresponding to the power-law spectrum
for different $\alpha$ values, ranging between 2 and -2.  
It is evident that the sources in the WBS are in agreement with the path described by the power-law model,
confirming the non-thermal origin of their infrared emission.
{\bf There is also a marginal evidence that HBLs tend to lie below the power-law line model
in Figure~\ref{fig:ints_bzb} while LBLs lie above the same. This deviation could be due to a contribution of the host galaxy or
to a mild curvature of their infrared spectra in the \wse\ bands.
Such interesting effect will be investigated in more details in a forthcoming paper \citep{massaro11b}}

\begin{figure}[]
\includegraphics[height=8.cm,width=8.5cm,angle=0]{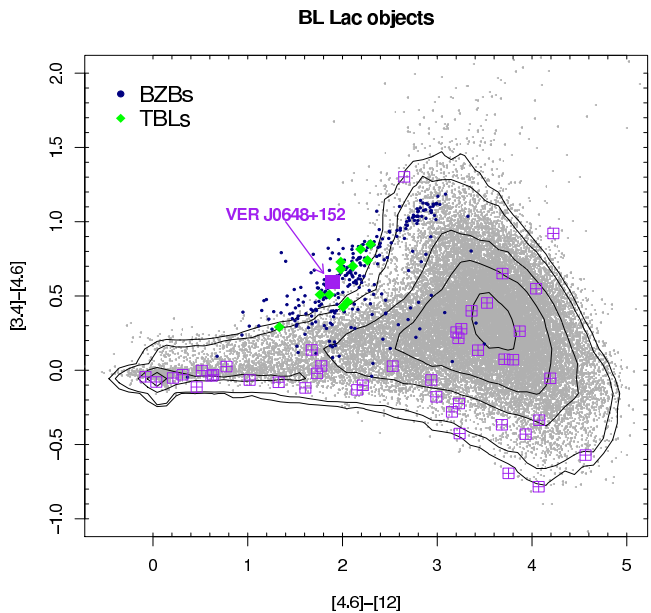}
\caption{Same of Figure~\ref{fig:ints_all}, only for the BL Lac objects (blue circles). 
The TBLs are highlighted (green diamonds). 
The purple open squares indicated all the sources associated to the \ver\ unidentified object
within the error box of the \fer\ position. The purple filled square points to the \wse\ source associated
with the X-ray position of XMM-Newton, also corresponding to the second closest source within 0.03 deg.}
\label{fig:ints_tev}
\end{figure}

Finally, we considered the list of HBLs detected at TeV energies (TBLs)
as reported in the TeVcat \footnote{http://tevcat.uchicago.edu/}, updated on 2011 August,
with $\Phi_{XR} >$ 1.
Only 13 TBLs are present in the sky mapped by \wse\ during the first year, and all of them have a counterpart
in the \wse\ archive within a region of radius 6$^{\prime\prime}$.5 while this number decreases to 11
if considering the smaller region of radius 2$^{\prime\prime}$.4.
We marked the TBLs in Figure~\ref{fig:ints_tev} in order to investigate if they have a 
different behavior from the TeV undetected HBLs (UBLs) as found in the X-ray analysis
of a subsample of the ROMA-BZCAT \citep{massaro11a}.
We found that the TBLs are more concentrated near the center of the WBS (see Figure~\ref{fig:ints_tev}).

\begin{figure}[]
\includegraphics[height=8.cm,width=8.5cm,angle=0]{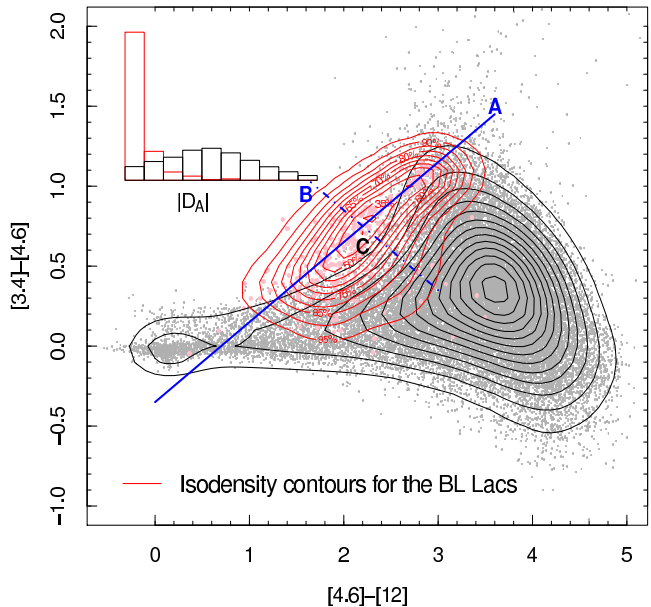}
\caption{The [3.4]-[4.6]-[12] $\mu$m color-color diagram for the BZB population and the \wse\ thermal sources
with the isodensity contours generated by KDE method. 
Lines A and B represent the ``axis" of the \wse\ strip occupied by BZBs and the directions of maximum 
gradient of the ratio between probability
density of the BZBs to the \wse\ thermal sources. The inset histogram shows the normalized distribution of 
distances of BZBs sources (red) and \wse\ thermal sources (black) calculated along the line A
(see Section~\ref{sec:plot}).
Labels on the contours indicate the percentage of BL Lacs contained in each contour.}
\label{fig:ints_kde}
\end{figure}

To highlight the separation between the WBS occupied by the BZBs in the [3.4]-[4.6]-[12] $\mu$m color-color 
diagram and the region spanned by the overall distribution of \wse\ extragalactic sources, we have evaluated the two-dimensional 
densities of both populations of source using the Kernel Density Estimation (KDE) technique (Figure~\ref{fig:ints_kde})
\citep[see e.g., ][and reference therein]{dabrusco09,laurino11}.
The KDE method provides an effective way of estimating the probability function of a multivariate
variable and do not require any assumption about the shape of the ``parent" distributions. In Figure~\ref{fig:ints_kde}, 
the isodensity contours drawn from the KDE density probabilities and associated with different levels of density 
are plotted for the two classes of sources (i.e., BZBs and \wse\ thermal sources).
The separation between the region of higher density of the BZBs distribution 
and that of the general \wse\ source distribution is evident. 

In particular, the line A shown in Figure~\ref{fig:ints_kde} has been obtained by a linear 
regression of the BZB density population in the [3.4]-[4.6]-[12] $\mu$m color-color diagram,
using the 95\% of BZBs detected by \wse\ in 2$^{\prime\prime}$.4.
It represents the main WBS axis. 
The second WBS axis (i.e., line B in Figure~\ref{fig:ints_kde})
displays the direction of maximum gradient of the density ratio for the distributions of the two populations
(i.e., BZBs and the \wse\ thermal sources).
The strong density gradient, visible along the line B, suggests a fairly good separation between the 
two distributions of BZBs and \wse\ thermal sources. 
The ratio of the BZBs density to the \wse\ thermal sources density increases
toward its maximum (i.e., point C in Figure~\ref{fig:ints_kde}) and along the curve A, 
while the density of the \wse\ thermal sources decreases.

The separation of the WBS from the distribution of \wse\ thermal sources is also evident in 
the normalized density histograms of the distances $|D_A|$ from the line A for both populations
(see inset of Figure~\ref{fig:ints_kde}). 
The peak of the BZB density histogram (inset of Figure~\ref{fig:ints_kde}) 
suggest a strong clustering of this source population, while that of the \wse\ thermal sources, 
with a completely different shape, peaks at a significantly large distance from the axis of the WBS . 

\section{VER J0648+152: an unidentified TeV source}
\label{sec:tevsource}
The blazar position over the WBS can be used as diagnostic tool to
identify extragalactic $\gamma$-ray sources.
We applied this to the case of the TeV unidentified source \ver\ recently 
discovered by VERITAS \citep{ong10},
that has been likely associated with a blazar at low galactic latitudes \citep{stephen10}.
The blazar identification has been suggested on the basis of both the
X-ray counterpart \citep{stephen10} and the \fer\ detection of 
the $\gamma$-ray source 1FGLJ0648.8+1516
within the positional error circle of 0.03 deg \citep{abdo10} .

We considered the positional circle of radius 108$^{\prime\prime}$ = 0.03 deg,
corresponding to the error on the position of the \fer\ source and we 
report the infrared colors for all the \wse\ sources over the [3.4]-[4.6]-[12] $\mu$m color-color diagram.
We found that the closest \wse\ object (20$^{\prime\prime}$) to the \fer\ position lies out of the WBS
in the area of the [3.4]-[4.6]-[12] $\mu$m color-color mostly dominated by stars and elliptical galaxies.
However, the second closest \wse\ source (29$^{\prime\prime}$) lie in the region of the WBS
where the TBLs are mostly concentrated (see Figure~\ref{fig:ints_tev}), 
corresponding to the position of the X-ray counterpart within the error box of XMM-Newton
(i.e., 7$^{\prime\prime}$.2) \citep{stephen10}.
In Figure~\ref{fig:ints_tev}, the open squares indicate all the sources 
associated with the \ver\ unidentified object
within the error box of the \fer\ position, while the filled square points to corresponding 
the \wse\ source.
{\bf The association of this \wse\ source strongly supports its blazar classification 
and given its position on the WBS, we suggest it is a new TBL.}

\section{Summary and Discussion}
\label{sec:discuss}
The most striking result of our investigation is that in the [3.4]-[4.6]-[12] $\mu$m color-color diagram
blazars lie in a distinct region (WBS) with respect to the other extragalactic sources.
This is due to the fact that blazars are dominated by non-thermal emission,
while all the other extragalactic classes of galaxies by thermal radiation at infrared frequencies.
This is strongly supported by the infrared colors of the WBS that are in agreement with a
power-law model for the IR spectrum of the BZBs (see also Figure~\ref{fig:ints_bzb}).

A similar attempt has been previously performed using the 2MASS archival data \citep[e.g.,][]{chen05},
to investigate the behavior of blazars with respect to the normal galaxies in the J-H-K color-color diagram.
However, our new approach has two main advantages:
first, because we are using mid-IR wavelengths where the stellar photospheric contributions
do not dominate galaxy colors,
the blazar population covers a distinct region in the [3.4]-[4.6]-[12]$\mu$m color-color plot, 
that separates them from the other extragalactic sources;
second, because the Galactic absorption at the \wse\ wavelengths is negligible.
 
The discovery of the WBS has several implications.
When the \wse\ data over the whole sky will be available it will be possible 
place all sources over the [3.4]-[4.6]-[12] $\mu$m color-color diagram
and those unidentified that lie in the WBS could be considered blazar candidates.
Multifrequency observations can be used to rule out this hypothesis;
this will be also simplified given the good astrometric precision of \wse\ data.

An additional use of the regions in the WBS is that they provide diagnostic tool for the blazar associations.
As previously discussed in Section~\ref{sec:detect}, in our analysis we excluded the 
multiple associations between a \wse\ source and a blazar in the ROMA-BZCAT.
However, our investigation shows that if at the position of the ROMA-BZCAT source
within few arcseconds there is a correspondent source in the \strip, this could be identified 
and associated with the blazar.

A samilar procedure could be applied to solve the problem of the unidentified $\gamma$-ray sources
(see Section~\ref{sec:tevsource}).
In the $\gamma$-ray sky most of the extragalactic $\gamma$-ray sources are blazars, 
in particular those that are emitting at GeV-TeV energies, where the HBLs dominate.
We showed the use of the WBS as diagnostic tool for the case of \ver\ recently detected at TeV energies 
by VERITAS \citep{ong10}, and already present in the 1st year \fer\ catalog \citep{abdo10}.
Our analysis of the infrared colors for all the \wse\ sources in the \fer\ positional error circle
strongly support its blazar classification \citep{stephen10}.
Its position on the WBS suggest that \ver\ is a new TBL.

Finally, we note that the region of the WBS covered by the TBLs is well defined.
Then, when the whole \wse\ data is released this region could be used to identify 
new TeV candidates for future observations.
A more accurate analysis on the possible use and developments of the WBS as
a diagnostic tool to classify blazars as well as 
the unidentified $\gamma$-ray sources will be presented
in more details in a forthcoming paper \citep{dabrusco11,massaro11b}.

\acknowledgements
We thank the anonymous referee for useful comments that led to improvements in the letter.
F. Massaro is grateful to M. Ashby, D. Harris, S. Willner for their helpful and fruitful discussions.
The work at SAO is supported in part by the NASA grant NNX10AD50G and NNX10AD68G.
R. D'Abrusco gratefully acknowledges the financial support of the US Virtual Astronomical Observatory, which is sponsored by the
National Science Foundation and the National Aeronautics and Space Administration.
F. Massaro acknowledges the Fondazione Angelo Della Riccia for the grant awarded him to support 
his research at SAO during 2011 and the Foundation BLANCEFLOR Boncompagni-Ludovisi, n'ee Bildt
for the grant awarded him in 2010 to support his research.
TOPCAT\footnote{\underline{http://www.star.bris.ac.uk/$\sim$mbt/topcat/}} 
\citep{taylor2005} was used extensively in this work for the preparation and manipulation of the tabular data.
Part of this work is based on archival data, software or on-line services provided by the ASI Science Data Center.
This publication makes use of data products from the Wide-field Infrared Survey Explorer, which is a joint project of the University of California, 
Los Angeles, and the Jet Propulsion Laboratory/California Institute of Technology, funded by the National Aeronautics and Space Administration.

{}

\end{document}